\documentclass[aps,pra,preprint,showpacs,amsmath,amssymb]{revtex4}
\input{epsf}

\usepackage{graphicx}
\usepackage{array}
\usepackage{bigstrut}
\usepackage{longtable}
\usepackage{rotating,booktabs}
\usepackage{booktabs,threeparttable}
\usepackage{color}
\usepackage{bm}

\begin{document}
\title{\bf{Observability of the Efimov spectrum in an electron-atom-atom system}}

\author{Huili Han}
\altaffiliation{Email: huilihan@wipm.ac.cn}
\affiliation {$^1$State Key Laboratory of Magnetic Resonance and Atomic and Molecular Physics, Wuhan Institute of Physics and Mathematics, Chinese Academy of Sciences, Wuhan 430071, China}

\author{Chris H.\ Greene}
\altaffiliation{Email: chgreene@purdue.edu}
\affiliation{Department of Physics and Astronomy,
Purdue University, 47907 West Lafayette, IN, USA}

\date{\today}

\begin{abstract}
The bound states of a system consisting of two heavy identical atoms and one light electron interacting through the finite-range pairwise potentials are explored, focusing on their dependence on the electron-atom scattering length. In the case of an exact resonance in the electron-atom interaction, the binding energy of an electron yields an effective $1/r^{2}$ potential for the relative motion of the atoms; One major finding is a universal potential that depends on the polarization length which goes beyond the Efimov region. An analytic expression for that potential is extracted from numerical calculations. The spectrum of the e+Rb+Rb system produced by the electron-atom polarization interaction exhibits three main sections, a non-universal transition region, a quasi-Efimov region, and a densely packed Efimov region.

\end{abstract}


\maketitle

\section{Introduction}
The Efimov effect is certainly among the most interesting and unexpected phenomena arising in few-body physics. It was pointed out by Efimov in 1970~\cite{Efimov,EFIMOV1973157}, and by now has been observed in many experiments with ultracold homonuclear Bose~\cite{NatEfimov,kraemer2006evidence,PhysRevLett.107.120401,PhysRevLett.103.163202}, three-component Fermi~\cite{PhysRevLett.103.130404,PhysRevLett.102.165302,PhysRevLett.101.203202}, and heteronuclear Bose-Fermi gases~\cite{PhysRevLett.115.043201,PhysRevLett.112.250404,PhysRevLett.113.240402,PhysRevLett.118.163401,PhysRevLett.117.163201}. Furthermore, the excited state of the helium trimer $^{4}$He$_{3}$ has recently been demonstrated to be an Efimov state~\cite{Kunitski551}.

In an atomic system, the Efimov effect occurs in a regime where the two-body scattering length $a$ is much larger than the characteristic
range r$_{0}$ of the interatomic interactions~\cite{Efimov}. In this regime, the Efimov states show a universal discrete scaling symmetry that is insensitive to the short range details of the two-body interactions. The three-body spectrum is invariant under a discrete scaling transform $E_{n}= \lambda^{2}E_{n+1}$, where $\lambda=e^{\frac{\pi}{s_{0}}}$ is the scaling constant and $s_{0}$ is a universal constant that depends only on the mass ratio,
the number of resonant pairs of interactions, and the identical particle symmetry~\cite{EricReview}. The Efimov effect is manifested in both bound and
scattering three-body observables that can be probed through observations of a series
of minima or maxima in the three-body collision processes such as three-body recombination, collision-induced dissociation, and vibrational relaxation~\cite{PhysRevA.72.032710,PhysRevA.73.030703}. Even granting the successes to date in confirming the Efimov effect experimentally, quantitative confirmation of the discrete scaling symmetry still remains a challenging goal that needs to be tested through the observation of multiple Efimov resonances. This challenge is acute in homonuclear systems with a large scaling constant $e^{\pi/s_0} \approx  22.7 $~\cite{EricReview,NIELSEN2001373}. Heteronuclear systems consisting of one light atom resonantly interacting with two heavy atoms, however, have a scaling constant significantly smaller than $22.7$, and thus have a significantly denser level spectrum. With this in mind, the Efimov effect has been investigated in K-Rb~\cite{PhysRevLett.111.105301, PhysRevLett.103.043201, PhysRevLett.118.163401, PhysRevLett.117.163201}, Rb-Li~\cite{PhysRevLett.115.043201} and Cs-Li~\cite{PhysRevLett.112.250404,PhysRevLett.113.240402} heteronuclear systems. Most notably, in the recent Li-Cs mixture experiments, three consecutive Efimov loss resonances were observed in the measurement of three-body loss coefficients~\cite{PhysRevLett.112.250404,PhysRevLett.113.240402}, and these have finally allowed an exploration of the universal scaling factors. The most favourable case for the appearance of the Efimov series of resonant features occurs when the mass ratio $m_{L}/m_{H}$ approaches $0$, leading to the scaling factor $\lambda\rightarrow 1$~\cite{EricReview}. The present paper explores an extreme realization of such a three-particle system, namely two identical bosonic atoms and an electron. The electron-atom interaction is dominated at long range by the charge-induced dipole moment interaction $- C_{4}/r^{4}$, where $C_{4} = \alpha/2$($\alpha$ is the neutral atom polarizability).  As the longest range term in the Hamiltonian, it plays the main role in determining the properties of the spectrum.

Extensive experimental and theoretical studies have now conclusively shown that a Van der Waals universality in the Efimov physics at large scattering lengths controls the properties of the lowest Efimov state in a system of three neutral atoms, whose long range potential $- C_{6}/r^{6}$ type interaction~\cite{PhysRevLett.109.243201,PhysRevLett.112.105301,PhysRevLett.111.053202,PhysRevA.90.043636}. However, the universal properties of such systems near the unitary limit that have $- C_{4}/r^{4}$ pairwise potentials is less well understood. Here we investigate the spectrum and explore the observability of the Efimov effect for a system of two atoms and an electron, with a near-resonant electron-atom interaction. Moreover, recall that in the framework of the Thomas-Fermi approximation, the electron-atom scattering length is a roughly periodic function of the atomic number~\cite{PhysRev.117.1281}. This means that a case of resonant electron-atom interaction might exist for some atom, whereby an Efimov effect might naturally occur for some negatively charged molecular system.  Alternatively, given the rapid improvements in quantum control capabilities in recent years, a resonant scattering length could conceivably be engineered through the application of electromagnetic field dressing to the system.

Several papers have dealt with exotic states of two neutral atoms and an electron~\cite{PhysRevA.60.3756,Pekov1,FONSECA1979273,pekov2}. The form of the effective potential for the electron-atom-atom system was first obtained using a Yamaguchi potential to represent the electron-atom interaction, which contains long-range components of an attractive Efimov-type potential $\propto 1/R^2$ ~\cite{Pekov1,pekov2}. Specifically, based on experimentally-known scattering phaseshifts for electron-helium scattering, suggested that even in the absence of any atom-atom interaction, a bound state of the He$_{2}^{-}$ system can exist~\cite{Pekov1}. Moreover, the lifetimes of the Efimov states for negative molecular ions were studied, and analytical expressions for widths and positions of resonances were derived, based on separable pairwise potentials~\cite{PhysRevA.60.3756}. In this paper, we investigate the spectrum properties and possible Efimov effects of two atoms and an electron system with resonantly electron-atom interaction. We use the finite-range pairwise potential to model the electron-atom and atom-atom interaction, which account for both the finite-range and polarization tail effects. This allows us to extract details about the universality associated with the polarization interaction.

This paper is organized as follows. In Sec.II, our theoretical method is presented. In Sec.III, we show the effective potentials for different polarization lengths. Properties for the potentials in different regions are discussed. The spectrum for the e-Rb-Rb system is also presented in Sec.III. Finally, the paper concludes with a brief summary and outlook.

\section{Theoretical method}
\begin{figure}
\includegraphics[width=0.65\textwidth]{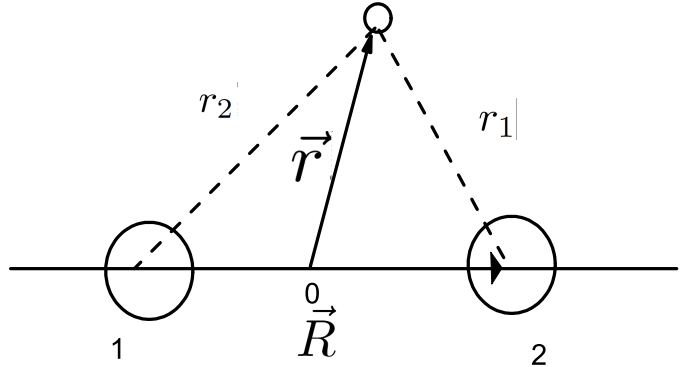}
\caption{The coordinates of the three particles used in the Born-Oppenheimer equation.} \label{f1}
\end{figure}
We consider the problem of two identical atoms of mass M and an electron interacting with each other by means of short-range potentials. The system is shown in Fig.~\ref{f1}. The wave function for the relative motion of system is determined by the three-body schr\"{o}dinger equation as follow(in a.u.)
\begin{equation}
\bigg[-\frac{1}{M} \nabla _{R}^{2} - \frac{1 }{2} \nabla_{r}^{2} + V_{AA}(R) + V_{eA}(r_{1}) + V_{eA}(r_{2}  ) \bigg] \Psi(r,R) = E \Psi(r,R)\\,
\label{eq1}
\end{equation}
where $R$ is the displacement vector between the two atoms with mass M, and $r$ is the vector from the center of mass of the atoms to the electron. In general, Eq.~$(\ref{eq1})$ can be solved within the hyperspherical framework. The hyperspherical potential curve picture reveals an intuitive way of understanding the bound and scattering properties of the system. However, for such molecular ion system, the hyperspherical method have the numerical convergence problem and the Born-Oppenheimer approximation is a better choice. This amounts to assuming that the wave function takes the form

\begin{equation}
\Psi(r,R)=\psi(r,R)\Phi(R)\,,
\label{eq2}
\end{equation}
where $\psi(r,R)$ is the wave function describing the motion of the electron when the two atoms have a fixed separation $R$. The electronic Schr\"{o}dinger equation separates into a pair of equations in prolate spheroidal coordinates. The first is the electron equation
\begin{equation}
\bigg[ - \frac{1 }{2} \nabla_{r}^{2} +  V_{eA}(\vec{R} - \vec{r}/2) + V_{eA}(\vec{R} + \vec{r}/2  ) \bigg] \psi(r,R) = \epsilon(R)\psi(r,R)
\label{eq3}
\end{equation}
where $\epsilon(R)$ is the energy eigenvalue depending parametrically on $R$. The equation for $\Phi(R)$ becomes
\begin{equation}
\bigg[-\frac{1}{M} \nabla _{R}^{2} + V_{AA}(R) + \epsilon(R)  \bigg] \Phi(R) = E \Phi(R)\,,
\label{eq4}
\end{equation}
We refer to this as the atom equation. The $\epsilon(R)$ plays the role of an effective potential in this equation. It is well known that the asymptotic behavior of $\epsilon(R)$ for this heavy-heavy-light system is
\begin{equation}
\epsilon(R)\approx - \frac{0.56714^{2}}{2 \mu R^{2}}\,,(R\ll a)
\label{eq5}
\end{equation}
 in the limit $|a| \rightarrow\infty$ ($\mu$ is the reduced mass of heavy-light atoms, $a$ is the scattering length of heavy-light atoms)~\cite{FONSECA1979273,0295-5075-101-6-60009,Braaten2006259}. This is the origin of Efimov physics, with a universal Efimov coefficient that controls the large-$R$ potential that is given by $s_{0}^{2} = 0.567143^{2} M/(2m) -1/4 $( where $m$ is the mass of the light particle). And in the limit $R\gg a$, the large-$R$ asymptotic form of $\epsilon(R)$ is:
\begin{equation}
\epsilon(R)\approx -\frac{1}{2 \mu a^{2}} - \frac{e^{-R/a }}{\mu aR} + \frac{e^{-2R/a}}{\mu aR}(1 - a/2 R)\,.
\end{equation}
This means that at large internuclear distances the exchange of the light atom leads to a Yukawa-type force between the two-heavy atoms~\cite{FONSECA1979273}.

In this paper, we employ elliptic (or prolate spheroidal) coordinates to solve Eq.~$(\ref{eq3})$ with B-splines as basis function. In these coordinates, the $\nabla_{r}^{2}$ operator is written as :
\begin{equation}
\nabla_{r}^{2}=\frac{4}{R^{2}(\xi^{2}-\eta^{2})}\bigg[\frac{\partial}{\partial \xi}(\xi^{2}-1)\frac{\partial}{\partial \xi} + \frac{\partial }{\partial \eta}(1-\eta^{2})\frac{\partial}{\partial \eta} + \frac{\xi^{2}-\eta^{2}}{(\xi^{2}-1)(1-\eta^{2})}\frac{\partial^{2}}{\partial \phi^{2}}\bigg]\,,
\end{equation}
 with
\begin{equation}
\xi=\frac{r_{1} + r_{2}}{R}\,, \eta=\frac{r_{1}-r_{2}}{R}\,.
\end{equation}
The ranges for ($\phi,\xi,\eta$) are: $0\leq \phi\leq 2 \pi$, $1\leq \xi\leq\infty$ and $-1\leq \eta\leq 1$. Since the wave function $\psi(\xi,\eta,\phi)$ is symmetric in the $\eta$ direction, we just shrink the integral range to $[-1,0]$ in the $\eta$ direction with boundary condition : $\frac{\partial \psi(\xi,\eta,\phi)}{\partial\eta}\bigg|_{\eta=0}=0$, which is appropriate since the two heavy particles are assumed to be identical bosons. The wave function for electron is expanded as a linear combination of B-spline basis functions in the $\xi$ and $\eta$ coordinates as follows:
\begin{equation}
\psi(\xi,\eta,\phi) = \sum_{n l}\frac{1}{\sqrt{2\pi}}B_{n}(\xi)B_{l}(\eta)e^{i m \phi}\,,
\end{equation}
 and $m$ is magnetic quantum number. In this paper, we consider only zero angular momentum and therefore set $m=0$ in our calculations.

As is common in studies of universality, we adopt soft-core potentials with the appropriate long-range behaviors to model the electron-atom and atom-atom interactions:
\begin{equation}
V(x)=-\frac{C_{n}}{x^{n}}[1-e^{-(\frac{x}{x_{c,n}})^{8}}]\\,
\end{equation}
where $n=4$ is for electron-atom interaction $V_{eA}(r_{1})$ and $n=6$ is for atom-atom interaction $V_{AA}(R)$. We tune the electron-atom and atom-atom interactions by directly altering the value of $x_{c,n}$. Consequently, changing the values of $x_{c,n}$ can be calibrated to produce the desired changes in electron-atom scattering length $a_{eA}$ and atom-atom scattering length $a_{AA}$. Analogous to the case of the van der Waals interaction, we define the polarization length $R_{\alpha}= \sqrt{2\mu C_{4}}/2$ (in a.u.) and the polarization energy as $E_{\alpha} = 1/(2\mu R_{\alpha}^{2})$ a.u., where $\mu$ is the two-body reduced mass. Typically, for electron-alkali atom interactions, these values are of order $R_{\alpha}\sim 10 $ a.u. and $ E_{\alpha}\sim 0.05$ a.u., respectively.

\section{Results and discussion}
\subsection{properties of effective potential}
\begin{figure}
\includegraphics[width=0.8\textwidth]{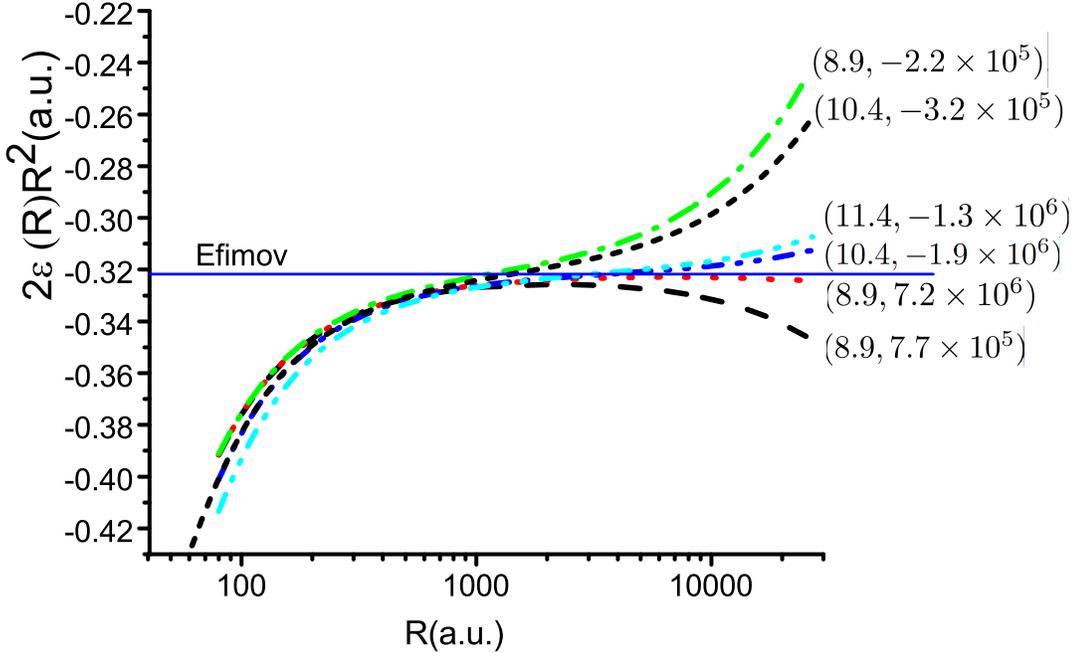}
\caption{Several rescaled potential curves $2 R^2 \epsilon(R)$ are shown as functions of $R$ for different polarization lengths and electron-atom scattering lengths $(R_{\alpha}, a_{eA})$.} \label{f2}
\end{figure}

\begin{figure}
\includegraphics[width=0.6\textwidth]{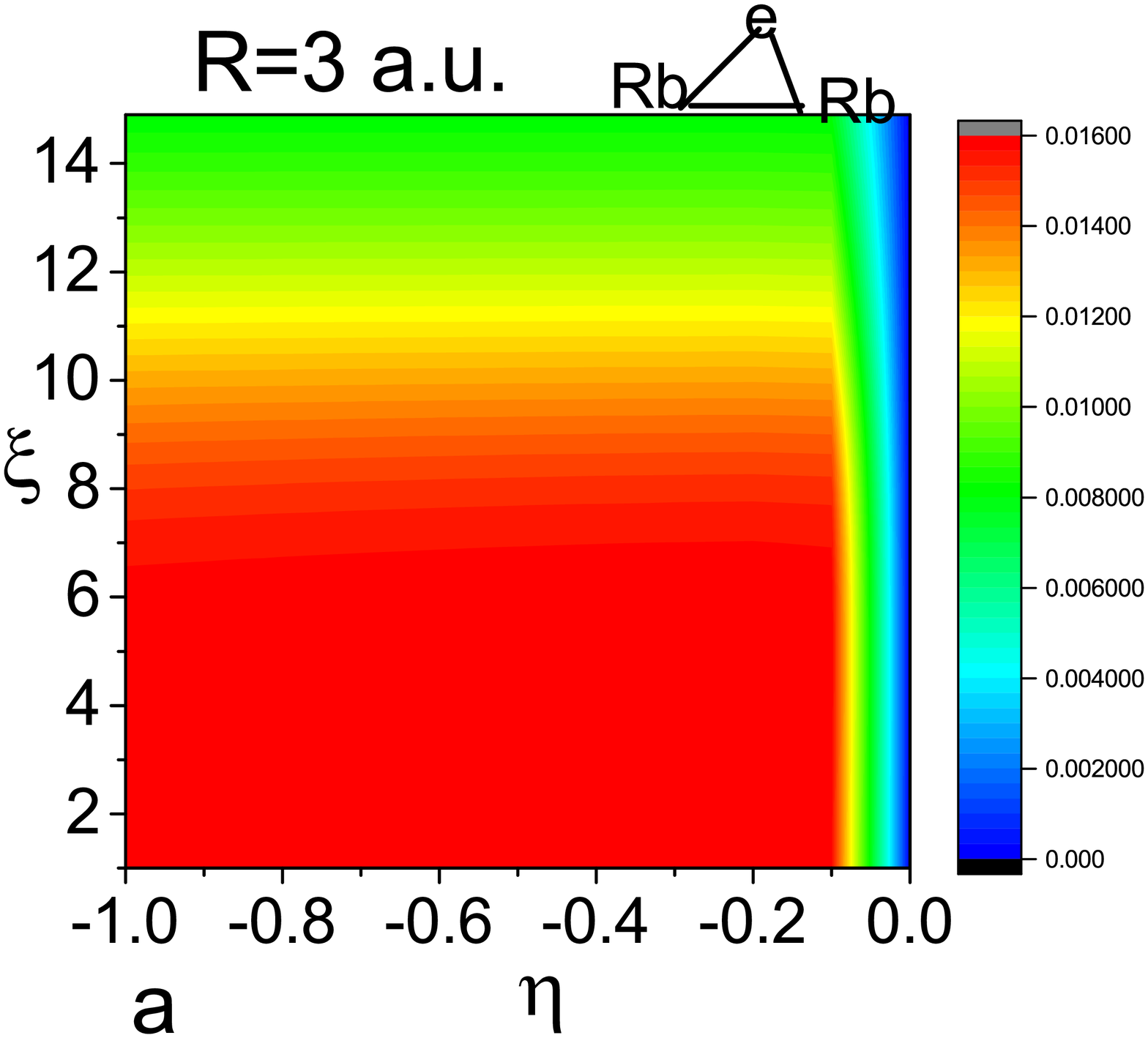}
\includegraphics[width=0.6\textwidth]{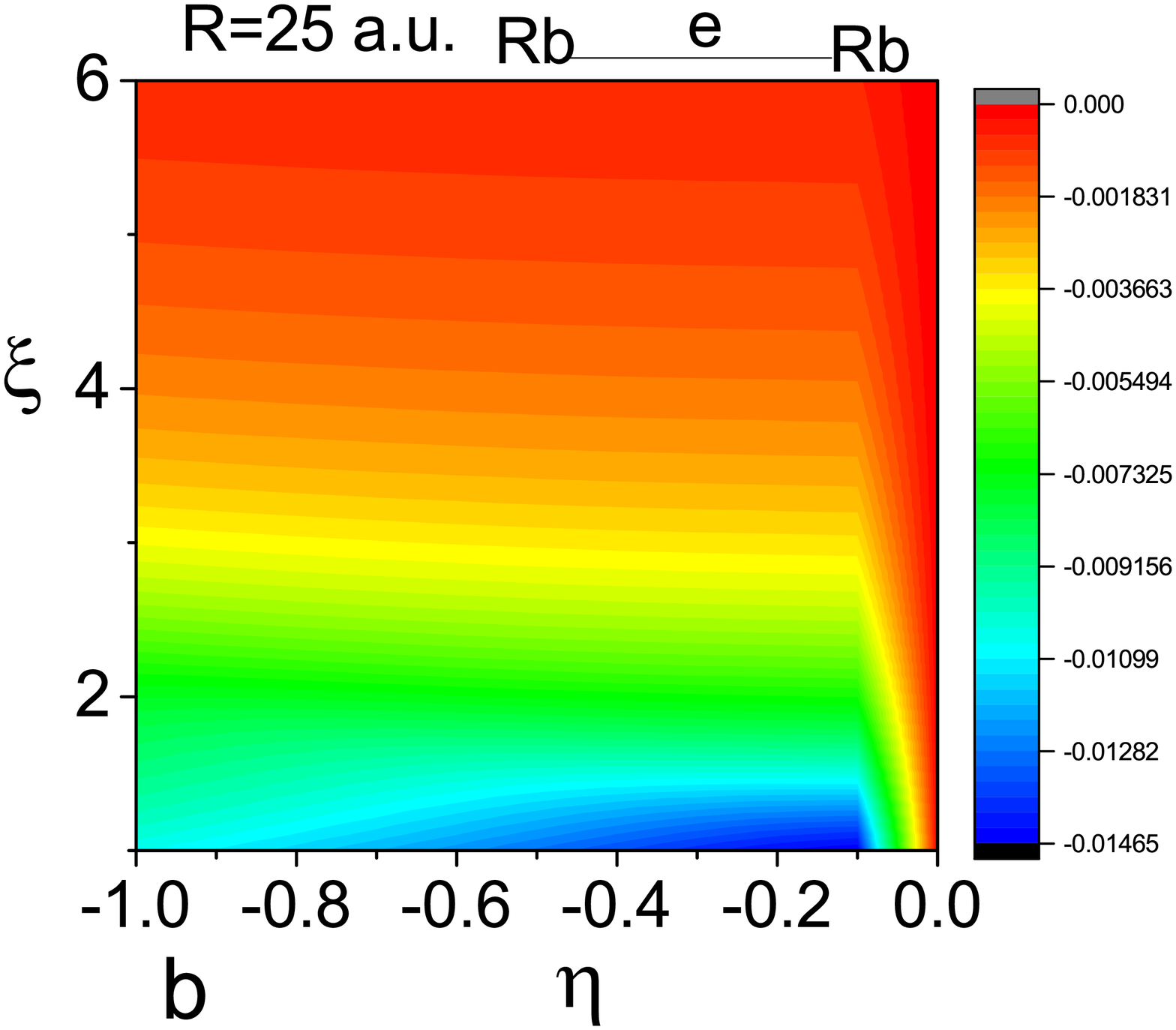}
\includegraphics[width=0.6\textwidth]{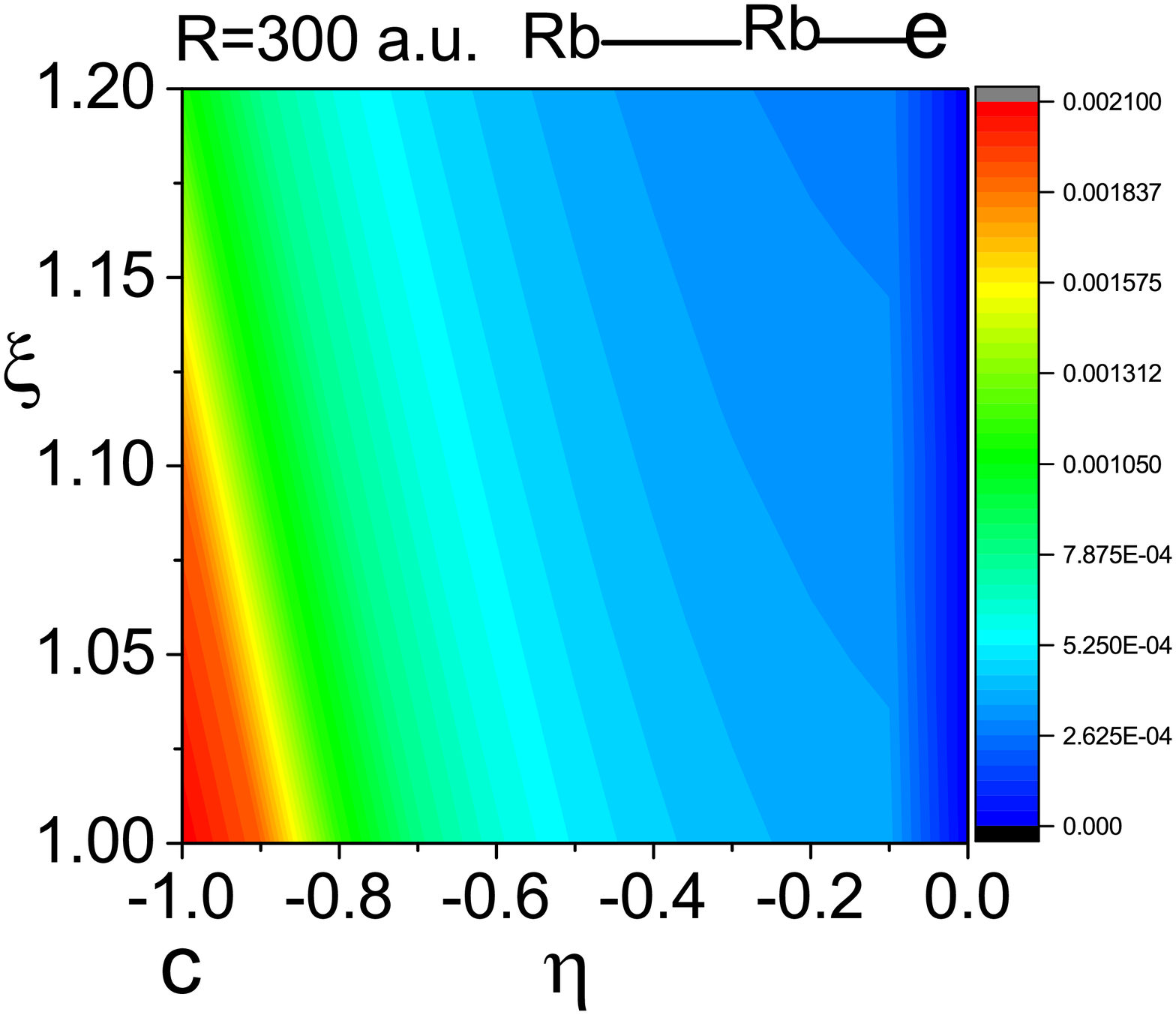}
\caption{Channel functions of the e-Rb-Rb system are plotted in spheroidal coordinates at three fixed values of $R$ in the universal limit. a) A channel function in the short range region; b) A channel function in the transition region; c) A channel function in the Efimov region.} \label{f3}
\end{figure}

The three-body effective potential, otherwise referred to as the Born-Oppenheimer potential energy curve, is $V_{eff}(R) = V_{AA}(R) + \varepsilon(R)$. For our numerical study, we consider a model with two identical atoms and one electron in the universal limit. To investigate the universal features of the effective potential, we plot the $\varepsilon(R)$ multiplied by $2R^2$ for different values of the polarization length $R_{\alpha}$, with the electron-atom scattering lengths very close to unitarity. The results are shown in Fig.~$\ref{f2}$. The graph shows that the curves have different features in the different ranges. We thus divide the effective potential into distinct regions in $R$: a short-range region, a transition region, a quasi-Efimov region and an Efimov region. We select the e-Rb-Rb system as an example to analyze the behavior in these different regions. For the e-Rb-Rb system, $s_{0}=159.621$ and the scaling constant is $\lambda=1.01988$; The polarizability for ground state of Rb is $319.2$ a.u.~\cite{0953-4075-43-20-202001}, giving a polarization length $R_{\alpha} = 8.9 $ a.u..  Figure~$\ref{f3}$ shows the channel function and structure of the e-Rb-Rb system at several fixed values of $R$ that are selected from the different regions.

For distance smaller than $R_{\alpha}$, the effective potentials depend on the short range details of the two-body interaction and $V_{AA}(R)$. For this reason, the potentials are non-universal in this region and will vary from system to system. We will not attempt to characterize the potentials in this region. Figure~$\ref{f3}a$ shows that the e-Rb-Rb three-body wavefunction has a triangular structure; the channel function is rather diffuse and not localized in this region.

\begin{figure}
\includegraphics[width=0.8\textwidth]{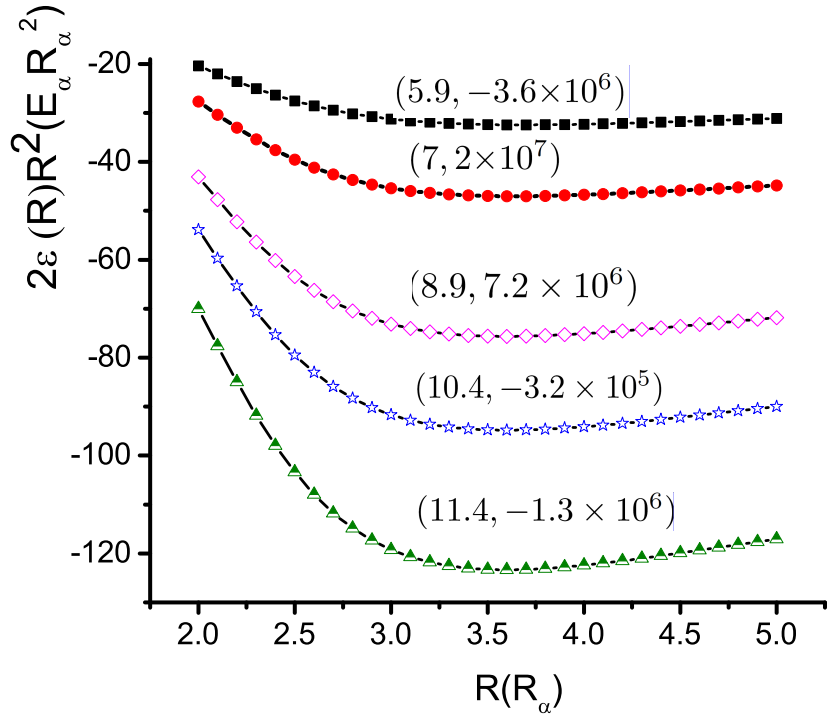}
\caption{$2\epsilon(R)R^2$ is plotted (in units of $E_\alpha R_\alpha^2$)
versus $R$ for different polarization lengths and electron-atom
scattering lengths $(R_{\alpha}, a_{eA})$ in the transition region.
} \label{f4}
\end{figure}

Figure~$\ref{f4}$ shows the rescaled potential $\epsilon(R)$ as function of R for different polarization lengths and electron atom scattering lengths $(R_{\alpha}, a_{eA})$ in the $R_{\alpha}<R<5R_{\alpha}$ range. All the curves with different polarization lengths show a rather flat region in this range, meaning that the effective potentials behave roughly like $-\gamma/R^{2}$ in this range. However, the values of $\gamma$ do not correspond to the Efimov case. The values of $\gamma$ are determined by the scattering length $a_{eA}$, and can be larger or smaller than the universal value of the Efimov coefficient. The flat region is universal when $R$ is scaled with the $R_{\alpha}$, which always begins from about $3 R_{\alpha}$. In this range, the structure of the three-body e-Rb-Rb system differs significantly from the short range behavior. As shown in Fig.~$\ref{f3}b$, the electron stays between the two atoms and the system approximately acquires a linear structure. We denote the range $R_{\alpha}<R<5R_{\alpha}$ as the "transition region".

 The Efimov region begins from about $R>30 R_{\alpha}$ for all of the curves shown, meaning that the a form of universality in the three-body parameter also holds for electron-atom-atom systems. The analytic form of $\epsilon(R)$ for this region is well known from the zero-range approximation. However, the approximation does not account for the region beyond the Efimov region, where the properties of $\epsilon(R)$ are not known before.

\begin{figure}
\includegraphics[width=0.8\textwidth]{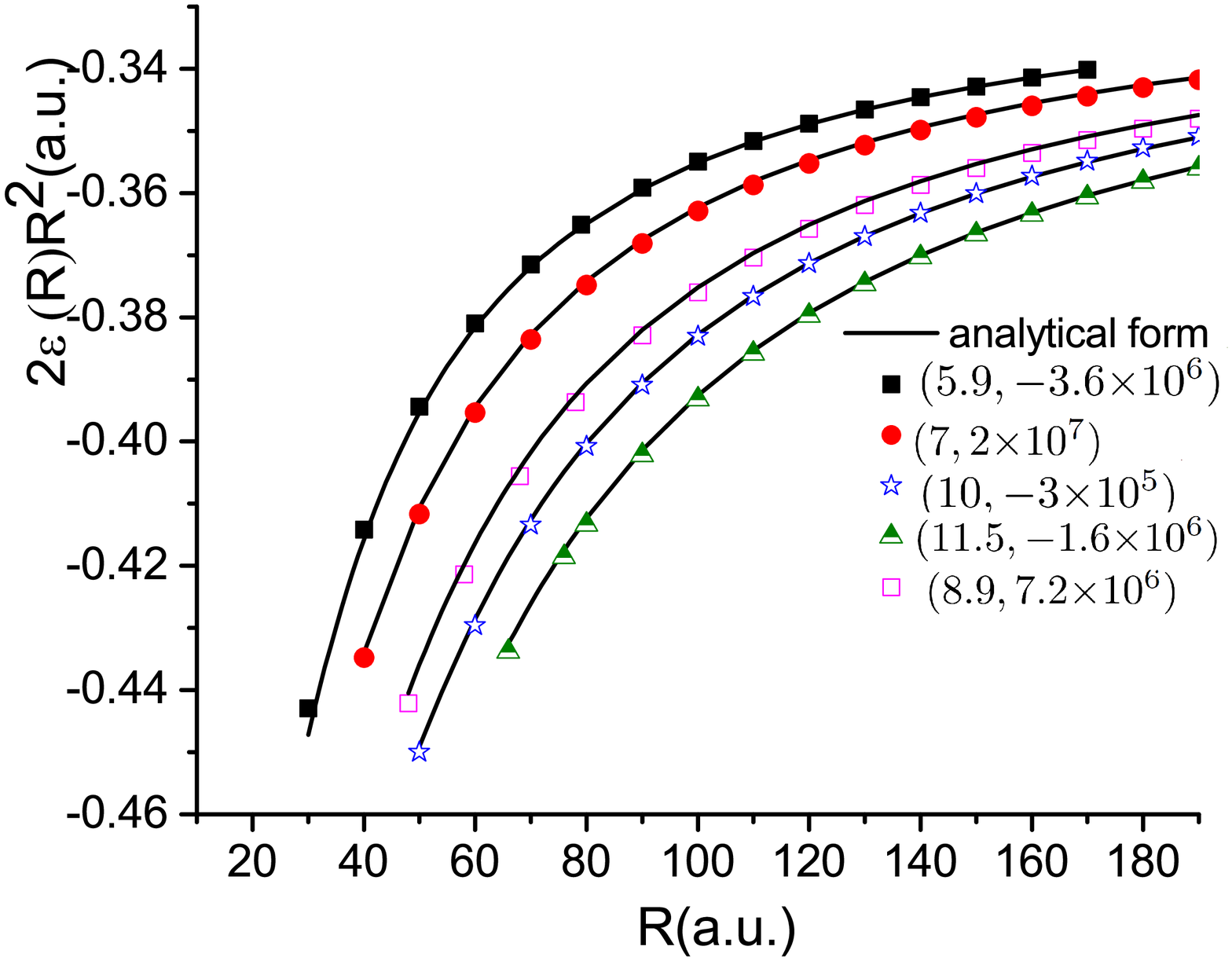}
\caption{Comparison between the analytic form of $\epsilon(R)$ and our numerical calculations at different values of the polarization length and electron-atom scattering length $(R_{\alpha}, a_{eA})$.} \label{f5}
\end{figure}

We focus on the $5R_{\alpha}<R<30R_{\alpha}$ range which is outside of the short-range region of $R$, but not yet out to the Efimov region. Examining Fig.~$\ref{f2}$, we can see that just before the Efimov region, the curves are controlled mainly by the value of $R_{\alpha}$. Thus a universal part of effective potential has apparently been found numerically, where the potential depends on the polarization length only. The analytic form of $\epsilon(R)$ in this region is :
\begin{equation}
\epsilon(R)\sim -\frac{0.56714^{2}}{2 R^{2}} - 0.25 \frac{R_{\alpha}}{R^{3}} - \frac{R_{\alpha}^{2}}{R^{4}}\,,
\end{equation}
which is obtained through fitting our numerical values. We call the range $5R_{\alpha}<R<30R_{\alpha}$ the quasi-Efimov region. Fig~$\ref{f5}$ shows the comparison between the analytic form $R^{2}\epsilon(R)$ and our numerical calculations for different $R_{\alpha}$ at unitary case. The figure shows good agreement between the analytic form and the numerical calculations. In this region, the structure of three-body is like that of Efimov states: the electron has an enhanced probability to stay near one atom, and the system has a linear structure as is shown in the Fig.~$\ref{f3}c$.

\subsection{Spectrum of system}
Having elucidated the physical characteristics of the effective potential energy curve, we now investigate the structure of the energy eigenspectrum. As expected from the Efimov estimation formula, the number of Efimov states is governed by the mass ratio and the ratio between the electron-atom scattering length and the potential range, through the formula
\begin{equation}
N=\frac{|s_{0}|}{\pi}\ln\frac{|a_{eA}|}{r_{0}}.
\end{equation}
 However, this formular is insufficient to reveal the finer features of the spectrum. It is well known that the spectrum of Efimov states possesses a characteristic geometric scaling. This property permits one to find the whole spectrum if the energy of one level is known. In fact, for the electron-atom-atom system and presumably other systems with an extremely large $s_{0}$, the motion of the atoms is of quasiclassical nature and the effective potential has an extensive range where the potential deviates from the limiting $R^{-2}$ form. The resulting small deviations in energy and reciprocal scattering lengths produce a considerable change of the wave function phase, and as a result, the scaling property turns out to exist within a surprisingly narrow energy range~\cite{EFIMOV1973157}. It is natural to ask what is the nature of the spectrum beyond the strict Efimov scaling region. In this paper, we explore the e-Rb-Rb system as an example of this broader behavior.

\begin{figure}
\includegraphics[width=0.8\textwidth]{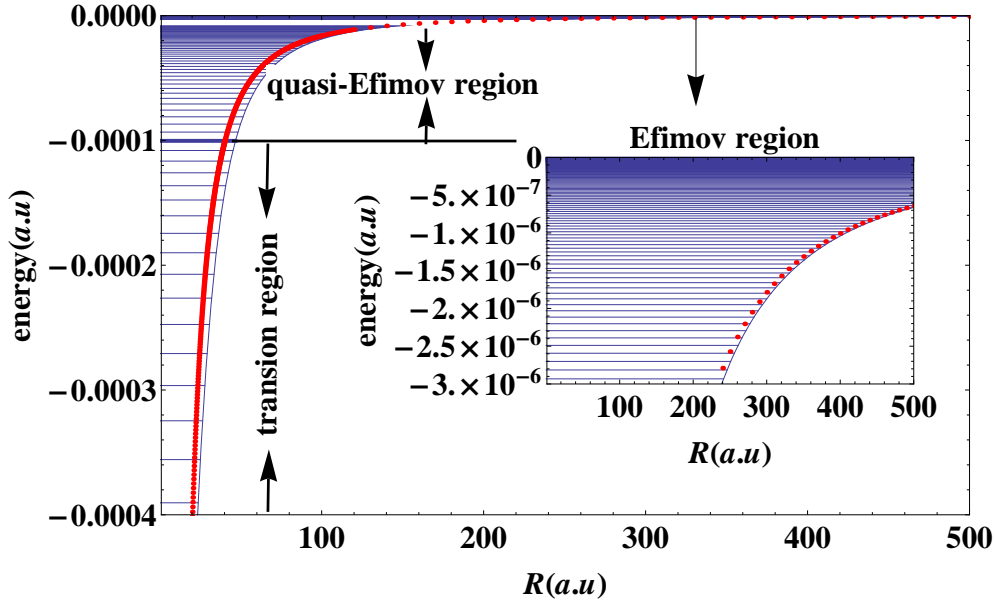}
\caption{Effective potential and $J=0$ spectrum for the e-Rb-Rb system at unitarity.} \label{f6}
\end{figure}

For the e-Rb-Rb system, the scaling constant is $\lambda=1.01988$, and consequently a very dense Efimov spectrum is expected. Figure~\ref{f6} shows the $J=0$ spectrum for the e-Rb-Rb system at unitarity, using a soft-core potential. The curve drawn as a collection of red points is the universal Efimov potential energy curve at unitarity for the e-Rb-Rb system, while the blue curve is our numerically computed effective potential. As expected, the system has a dense spectrum of energy levels. Analysis of the ratio of energy levels in the different regions shown in Fig.~\ref{f6} can help us to gain an intuitive understanding of the underlying physics and the deviations of the effective potential from the limiting Efimov case.

\begin{figure}
\includegraphics[width=0.6\textwidth]{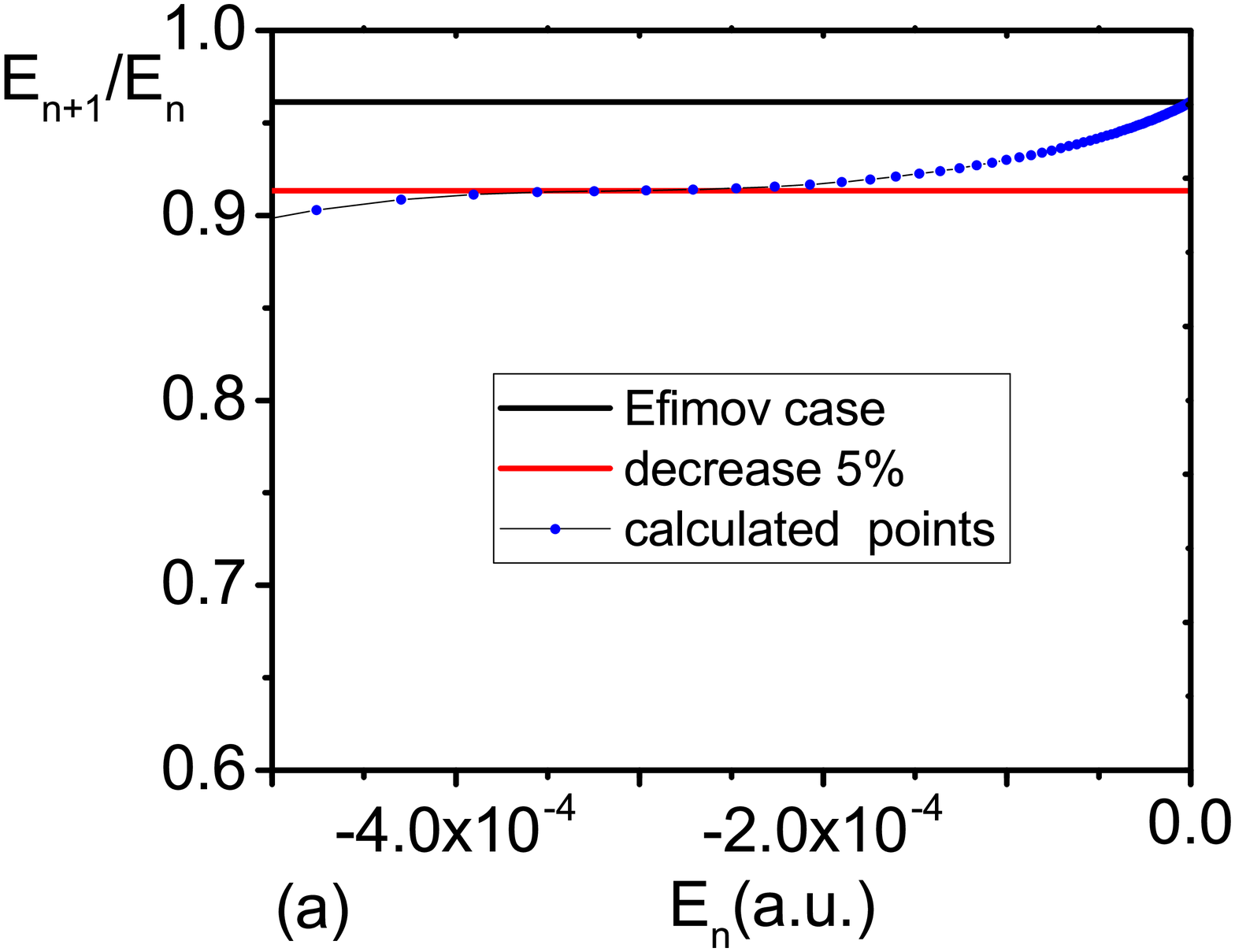}
\includegraphics[width=0.6\textwidth]{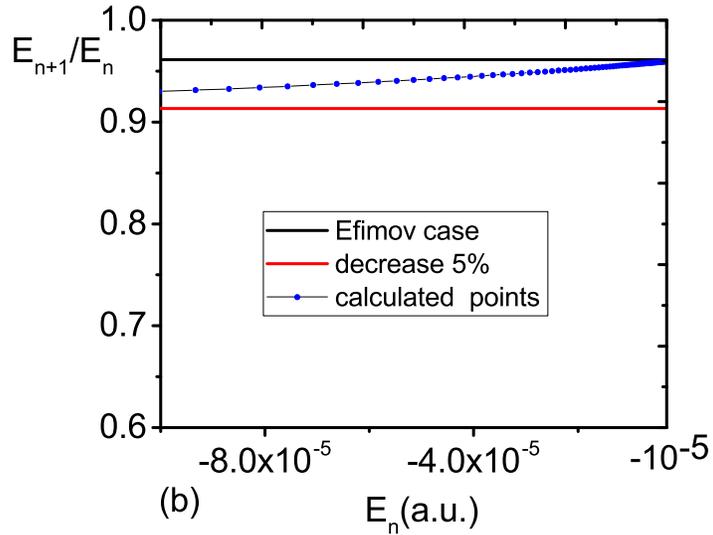}
\includegraphics[width=0.6\textwidth]{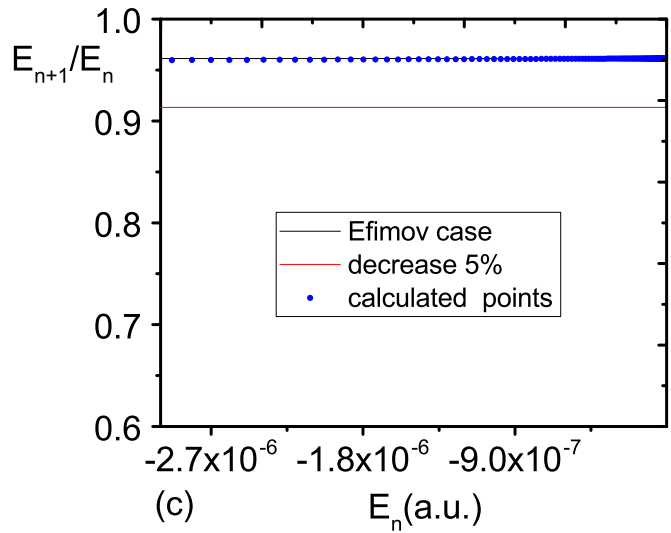}
\caption{Successive energy level ratios for the e-Rb-Rb system at unitarity in the three different regions: a) energy ratio for the transition region, b) energy ratio for the quasi-Efimov region, c) energy ration for the true Efimov region. The black solid line shows the energy ratio in the universal Efimov limit, while the red solid line is the value representing a $5\%$ deviation from the Efimov universality ratio. } \label{f7}
\end{figure}

Fig.~\ref{f6} shows that, in the transition region, the spectrum is far more sparse than in the higher regions closer to the dissociation threshold. Remarkably, the ratio of successive energy levels is approximately constant, with one approximately flat region apparent in the Fig.~\ref{f7}a. This means the supported potential locally in the vicinity of the turning points has an approximately $-\gamma/r^{2}$ form, which is consistent with the behavior of the effective potential in Fig.~\ref{f4}. The spectrum in the  quasi-Efimov region then become even more dense and the energy ratio exhibits its own approximate scaling in this region. Fig.~\ref{f7}b shows that this energy ratio deviates by only about $5\%$ from the expected universal Efimov ratio in this quasi-Efimov region. In the true Efimov region, energy levels accumulate at the three-body dissociate threshold and show the expected universal scaling properties. Fig.~\ref{f7}c shows the successive energy ratio in this region.

\begin{figure}
\includegraphics[width=0.8\textwidth]{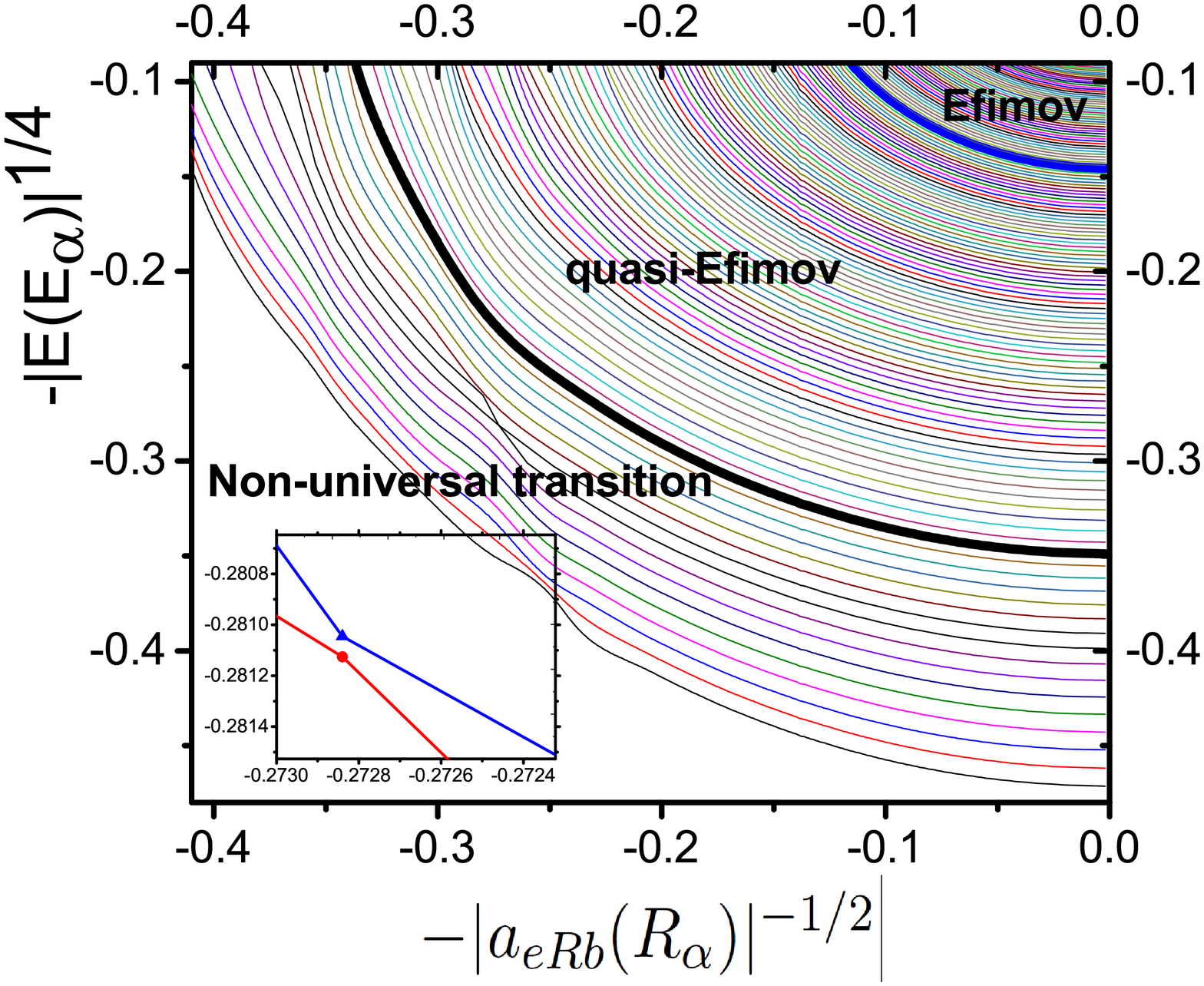}
\caption{Spectrum of the $J = 0$ e-Rb-Rb energy levels versus $-|a_{eRb}(R_{\alpha})|^{-1/2}$ with scattering length $a_{RbRb}$ fixed at the unitary limit. } \label{f8}
\end{figure}

Figure~\ref{f8} shows the spectrum of $J = 0$ levels for the e-Rb-Rb system obtained for a varying scattering length $a_{eRb}$ with $a_{RbRb}$ fixed at the unitary limit. In the Efimov region, the trajectories of energy levels are accurately parallel. The boundary of this region is given by the heavy blue line, which crosses the three-body dissociation threshold at $a_{eRb} = -486 a.u.$. Between the heavy black line and the heavy blue line is the quasi-Efimov region, where the level trajectories are roughly similar. In the transition region, all traces of similarity disappear and the ratios of successive energies are not even approximately constant. The deepest bound state of the system crosses the three-body dissociation threshold at $a_{eRb} = - 39 a.u.$.

\section{conclusion}
The present study employs the Born-Oppenheimer approximation to investigate the effective potential energy function in a three-body system consisting of two identical bosonic atoms and an electron. In the case of an exact resonance in the s-wave electron-atom interaction, the binding energy of an electron yields an effective $1/r^{2}$ potential for the relative motion of the atoms; Remarkably, part of the universal potential that depends on the polarization length is also found beyond the Efimov region. The analytic expression is given in an approximate form that reproduces our numerical values. Using the e-Rb-Rb system as an example, this investigation of the spectrum shows that a rich spectrum can exist even when there is no high-lying bound state of the two Rb atoms.

\section{Acknowledgments}
We thank P. Giannakeas for discussions in the early stages of this project.  This work was supported in part by the National Science Foundation Grant number PHY-1607180. Hui-Li Han was supported by National Key Research and Development Program of China under Grant No.2016YFA0301503 and National Natural Science Foundation of China under Grant No. 11634013.

 \end{document}